\documentclass[12pt]{article}\usepackage[hyperfootnotes=false]{hyperref}
\usepackage{epsfig}
\usepackage{float}
\usepackage{empheq}
\usepackage{bbold}

\usepackage{caption}

\usepackage{amsmath}
\usepackage{amssymb}
\usepackage{graphicx}
\newlength{\abstractwidth}
\renewcommand{\thefootnote}{\fnsymbol{footnote}}
\renewcommand{\thanks}[1]{\footnote{#1}} 
\newcommand{\starttext}{
\setcounter{footnote}{0}
\renewcommand{\thefootnote}{\arabic{footnote}}}

\newcommand{\be}{\begin{equation}}
\newcommand{\bea}{\begin{eqnarray}}
\newcommand{\eea}{\end{eqnarray}}
\newcommand{\beq}{\begin{equation}}
\newcommand{\ee}{\end{equation}}

\newcommand*\widefbox[1]{\fbox{\hspace{2em}#1\hspace{2em}}}

\def\ra{\rangle}
\def\simleq{\; \raise0.3ex\hbox{$<$\kern-0.75em
\raise-1.1ex\hbox{$\sim$}}\; }
\def\simgeq{\; \raise0.3ex\hbox{$>$\kern-0.75em
\raise-1.1ex\hbox{$\sim$}}\; }

\def\bi{\begin{itemize}}
\def\ei{\end{itemize}}

\def\sc{\setcounter{equation}{0}}

\def\CC{{\cal{C}}}

\def\bsub{ \begin{subequations}
\begin{empheq}[box=\widefbox]{align}  }
\def\esub{ \end{empheq}
\end{subequations}}

\def\1{\(  \mathbb{1} \)}

  \def\bn{\bigskip \noindent}

\makeatletter
\g@addto@macro\normalsize{%
  \setlength\abovedisplayskip{10pt}
  \setlength\belowdisplayskip{20pt}
  \setlength\abovedisplayshortskip{10pt}
  \setlength\belowdisplayshortskip{20pt}
}
\makeatother

\usepackage{color}

\begin{document}


  
\begin{titlepage}

\rightline{}
\bigskip
\bigskip\bigskip\bigskip\bigskip
\bigskip

\centerline{\Large \bf {Horizons Protect Church-Turing }}

\bn

\bigskip
\begin{center}
\bf   Leonard Susskind  \rm

\bigskip

SITP, Stanford University, Stanford, CA 94305, USA

Google, Mountain View CA, 94043, USA

\end{center}

\bn

\begin{abstract}

The quantum-Extended Church-Turing thesis is a principle  of physics as well as computer science. It asserts that the laws of physics will prevent the construction of a machine that can efficiently determine the results of any calculation which cannot be done efficiently by a quantum Turing machine (or a universal quantum circuit). In this note I will argue that an observer falling into a black hole can learn the result of such a calculation in a very short time, thereby seemingly violating the thesis. A viable reformulation requires that the thesis only applies to observers who have access to the holographic boundary of space. The properties of the horizon play a crucial a role in protecting the thesis. The arguments are closely related to, and were partially motivated by a recent paper by Bouland, Fefferman, and Vazirani, and by a question raised by Aaronson.

\end{abstract}

\end{titlepage}

\starttext \baselineskip=17.63pt \setcounter{footnote}{0}
\tableofcontents

\sc

\section{The Quantum-Extended Church-Turing Thesis}\label{Sec: QECT}
The original Church-Turing thesis may be regarded as a principle of physics. It states that any computation that can be done by a physical system can be done by a Turing machine. In its contrapositive form it says that a computation that cannot be done by a Turing machine cannot be done without violating a law of physics. It's a principle of physics because it implies a limitation on what physical systems can do. 
The CT thesis is similar to the principle that forbids perpetual motion machines: any claim of a perpetual motion machine will upon examination be found to violate some law of physics. Likewise, any claim that a machine can calculate what a Turing machine cannot, will upon examination be found to violate a law of physics. 

The Church-Turing thesis is believed to be correct.

The extended CT thesis (ECT thesis) goes further and says that any calculation that cannot be done \it efficiently \rm (in polynomial time) by a Turing machine, cannot be done efficiently by any physical system. 
The ETC thesis is widely  thought to be wrong; quantum machines are  able to  perform calculations in polynomial time that Turing machines are believed to not be able to do, e.g., factoring\footnote{It is widely believed but not proved that the factoring of large integers is not in the class P.}. 

This brings us to the quantum-Extended Church-Turing (qECT) thesis:
\bn

\it  Any calculation that cannot be done efficiently  by a quantum Turing machine (or quantum circuit), cannot be done efficiently by any physical system consistent with the laws of physics. \rm

\bn

The  quantum-Extended Church-Turing Thesis  (qECT thesis) has resisted all attempts to disprove it. But like anything in physics it may prove to be wrong when it is pushed beyond the limits of established laws. Will it survive the surprises that quantum gravity throws at us? 

Indeed black holes have the potential  to threaten to the qECT thesis.  It is computationally very difficult to determine what is going on behind the horizon of a black hole from knowledge of its quantum state \cite{Bouland:2019pvu}. In some cases it is exponentially difficult. But by jumping into the black hole  Alice   
can quickly gain access to  such information. 
This would  violate a naive version of the qECT thesis. 

In sections \ref{sec: C of C} and  \ref{sec: shocks}  I will set up a precise example, the important features of which I believe are general.  The example makes clear a quantum gravity formulation of the qECT thesis must be carefully formulated in order to be correct.
In section \ref{sec: violation} I will offer such a formulation.

\section{The  Complexity of  Complexity}\label{sec: C of C}

The computational complexity of a quantum state $|\Psi \ra $ is the minimal number of gates that are required to prepare   $|\Psi \ra $  from some simple reference state, for example a product state of the form $$|0\ra^{\otimes N}.$$  Quantum computational complexity can be exponentially large (in the entropy of the system), but in this paper we will only be interested in states of polynomial complexity. 

However there is another quantity which can grow exponentially large, even as the complexity remains polynomial.  I'll call it the ``complexity of complexity."  Let me explain: 
Suppose I give you a quantum state\footnote{By this we mean a polynomial number of replicas of the system.} of a chaotic quantum system and I tell you that its complexity is less than or equal to some number $\CC_0.$ How difficult is it to confirm that $\CC\leq \CC_0$  assuming you don't know the (possibly time dependent) Hamiltonian that created the state\footnote{If we know the Hamiltonian we can run it backward and see how long it takes to return to the reference state. This would place an upper bound on $\CC\CC$. In order to avoid this possibility Bouland, Fefferman, and Vazirani introduce an element of unknown randomness into the evolution in the form of a pseudorandom sequence of  shockwaves.  }? You'll need to do a computation which itself has some computational complexity. The complexity of complexity, which I will call $\CC\CC$, is the number of simple logical steps that it would take to confirm $\CC\leq \CC_0.$

Generically there are no shortcuts to computing computational complexity. One simply tries all circuits of $1$ gate, $2$ gates,
$3$ gates,....., until the target state  $|\Psi \ra $ is reached. Since the number of circuits with $n$ gates is exponential in $n$ it follows that the complexity of  complexity is exponential,
\be
\CC\CC \sim \exp{\CC_0.}
\ee

If we now apply the qECT thesis we must conclude that no physical process can   confirm that $\CC\leq \CC_0$  in time less than  $\exp{\CC_0}.$

The actual task that will interest us in this paper is slightly different than determining the complexity of  a state. We want to determine  whether the complexity is increasing or decreasing and if so, how fast. This can be done if we have copies of the state at two neighboring times. We expect that the complexity of determining the rate of change of complexity---call it $\CC\dot{\CC}$---is also exponential. Thus, if we accept the qECT thesis, then no physical system can determine whether complexity is increasing or decreasing in a time less than  $\exp{\CC_0}.$ 

In the next section I will show how states can be constructed for which complexity decreases with time, or more generally, grows at less than the normal rate. I will also give evidence that states of this type have anomalous behavior just behind the horizon---behavior that can be detected by an in-falling observer. Although from the outside  $\CC\dot{\CC}$  is exponentially large, an observer who crosses the horizon can quickly detect the anomaly and estimate the rate of complexification.

In  section \ref{sec: violation}   I will discuss the implications for  the qECT thesis. Naively it seems that the thesis is violated, but  I will argue that a correct version can be formulated, and that instead of violating the thesis, the black hole horizon provides the censorship which protects it. This suggests a new information-theoretic role for horizons.

\section{Shock Waves and the Rate of Complexification}\label{sec: shocks}

\begin{figure}[H]
\begin{center}
\includegraphics[scale=.2]{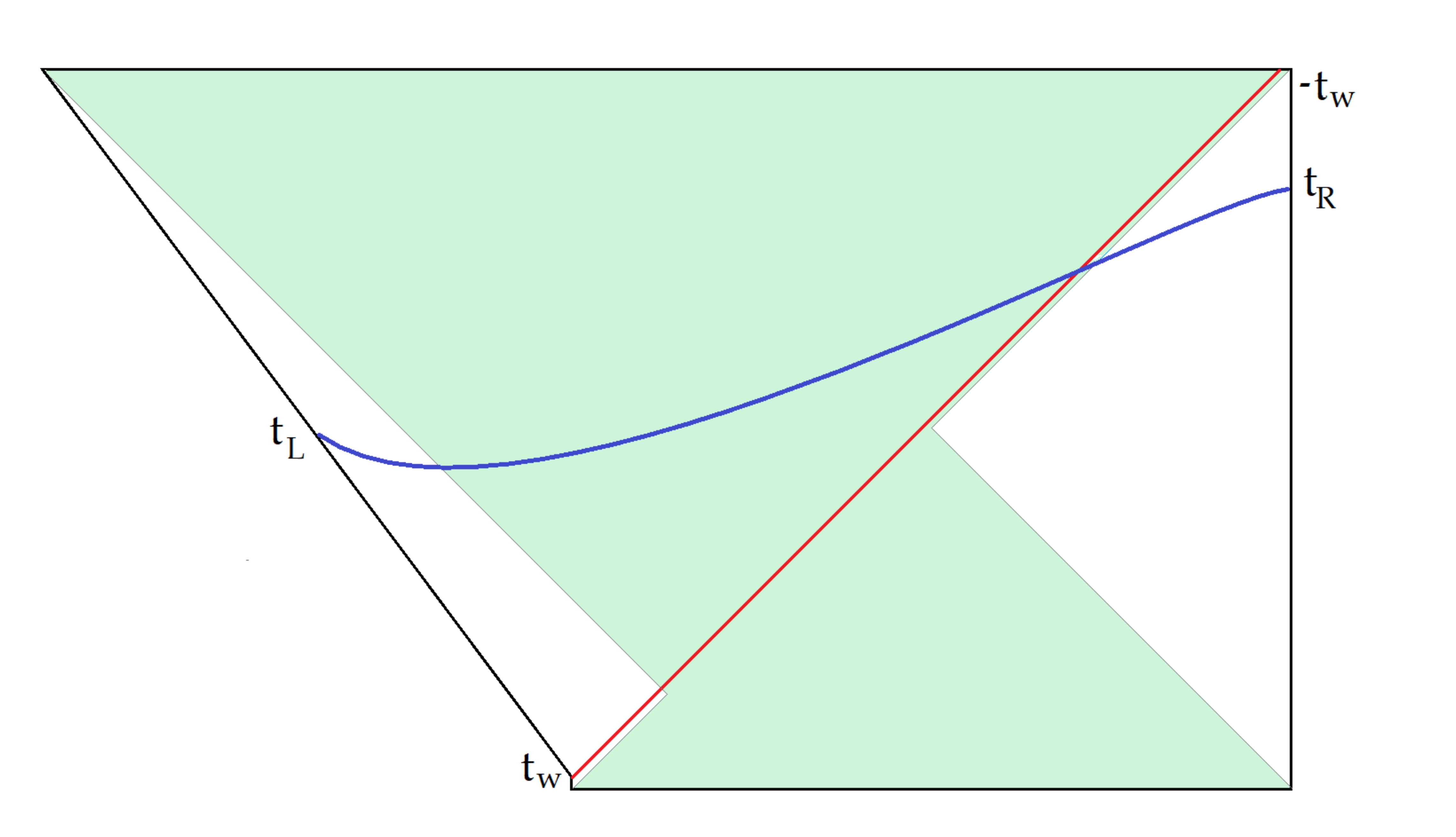}
\caption{ Penrose diagram for a two-sided black hole perturbed by a shock wave represented by the red line. The shock is generated by a low energy boundary  perturbation at 
$t_L = t_w.$ The blue curve represents a maximum-volume surface anchored at points $t_L, \  t_R.$   The green region is behind both horizons.  }
\label{Ct1-2}
\end{center}
\end{figure}

There is a simple  mechanism for illustrating how  decreasing complexity    \cite{Stanford:2014jda}\cite{Susskind:2015toa}\cite{Zhao:2017iul} can occur.
Consider a two-sided black hole in the thermofield double state. Bob lives on the left side and Alice on the right side. At some negative time $t_w$ in the past ($t$ is measured in units of the AdS radius of curvature) Bob applies the thermal scale perturbation $W$ as shown in (red) in figure  \ref{Ct1-2}.

We are interested in the complexity ($\CC,$ not $\CC\CC$) of the state as a function of the right-side time $t_R.$  According to the complexity-volume \cite{Brown:2015bva} conjecture this is proportional  to the volume of a maximal surface anchored at some fixed $t_L$ and some variable  right-side time $t_R.$ One finds that the  volume  satisfies \cite{Stanford:2014jda}\cite{Susskind:2015toa},
\be 
\exp{V(t_L, t_R)} \sim    \cosh \left(  \frac{t_L+t_R}{2} \right) + \frac{e^{|t_w| -t_* + \frac{t_L}{2}-\frac{t_R}{2} }}{2}   
\label{master}
\ee
where $t_*$ is the scrambling time \cite{Sekino:2008he}. We assume that $|t_w|>t_*.$

In order to eliminate dependence on $t_L$ let us assume that it is much larger than any of the other times in the problem,

\be 
t_L >>t_*, \  t_R, \   |t_w|
\ee

Define $\Delta = (|t_w| - t_*).$ We may write \ref{master},
\be 
{V(t_L, t_R)} =     {\frac{t_L +\Delta}{2}} +\log\left[ \cosh{\frac{t_R-\Delta}{2}} \right]
\ee
and
\be
\frac{dV}{d t_R} =\frac{1}{2} \tanh{\left(  \frac{t_R-\Delta}{2}  \right)}
\label{t-derivative}
\ee

We see:
\begin{enumerate}
\item The right hand side of \ref{t-derivative} is independent of $t_L.$
\item There is a fairly sharp transition from decreasing to increasing complexity at the  crossover time\footnote{Transition takes place over a time interval equal to the AdS radius of curvature. },
\be 
t_R^{crossover} = \Delta = |t_w| -t_*
\ee
\end{enumerate}

\begin{figure}[H]
\begin{center}
\includegraphics[scale=.25]{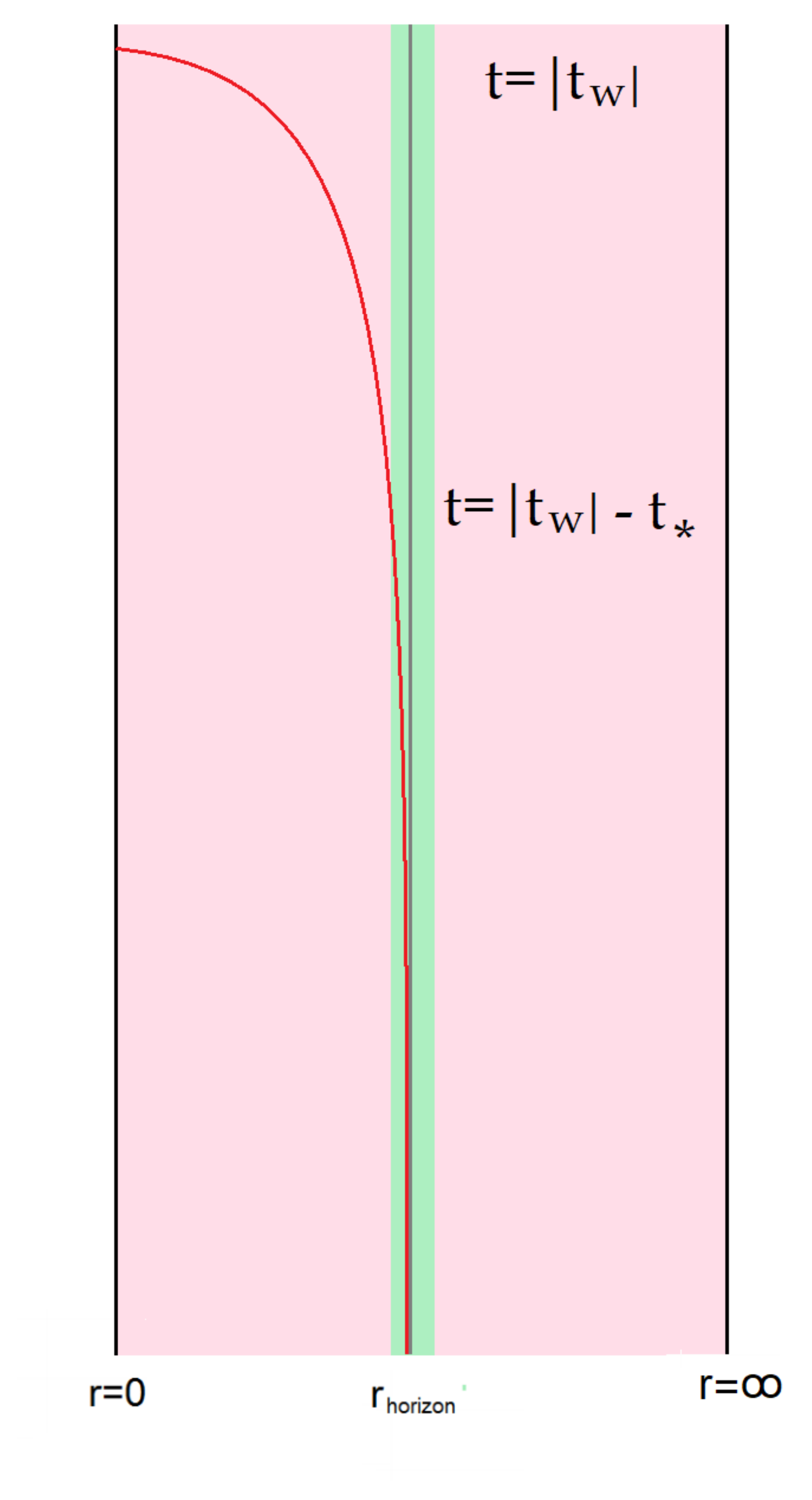}
\caption{ The same black hole as in figure \ref{Ct1-2} but in Eddington-Finkelstein coordinates. The horizontal time-slices are light-like. The green region is a generalized stretched horizon a Planck length thick. The red shock wave is initially in the stretched horizon but leaves the Planckian region at $t=\Delta = |t_w| -t_*.$ Subsequently it falls to the singularity at $t=|t_w|.$      }
\label{Ct2}
\end{center}
\end{figure}

To examine the properties of the crossover in more detail it is useful to redraw the picture in Edington-Finkelstein coordinates\footnote{The Edington-Finkelstein time coordinate is constant along in-falling light rays. At the AdS boundary it is equal to the usual boundary time.} as in figure \ref{Ct2}
In figure \ref{Ct2} the horizon is  surrounded by a   thickened ``stretched" horizon of Planckian width shown in green. The shock wave initially is contained within the stretched horizon, during which time its energy is superplanckian. An observer attempting to cross the horizon will be met by a very energetic firewall.

The crossover time $t= |t_w| - t_* = \Delta$ is precisely the point at which the shock wave leaves the Planckian stretched horizon. The shock wave then falls to the singularity at $r=0,$ reaching it at $t=|t_w|.$ During the time between $\Delta$ and $|t_w|$ the shock wave can be detected by an in-falling  observer, but as it falls it loses energy. By $|t_w|$ it has become a benign but detectable thermal quantum, and beyond that it is unobservable to an infalling observer. 

Over this period, between $\Delta$ and $|t_w|$, the complexity grows but at a rate slower than maximum. At $t=\Delta$ the rate is zero and by $t=|t_w|$ it has achieved its normal maximal value. 

To summarize: it is during the time that the complexity is evolving at a less-than-normal rate that a signature can be felt behind the horizon.
I will assume without further proof that this pattern is general.

\section{Violation of the qECT Thesis?} \label{sec: violation}

Aaronson\footnote{Scott Aaronson, private communication} has expressed skepticism
about the reasoning of the previous section. The argument goes as follows:

We expect that if a quantum circuit has evolved for a time $t,$ its complexity will be of order $t,$  and that the problem of determining whether complexity is increasing or decreasing will be exponentially complex,
\bea 
\CC &\sim& t \cr
\CC \dot{\CC} &\sim& \exp{t}.
\eea

 We expect this to also be true for black holes. The qECT thesis then implies  that the time required to made such a determination should also be exponential. But if the considerations of the previous section are correct,  Alice  can jump into the black hole and find out in  order $1$ time if the complexity is evolving in an abnormal way. This is  very surprising because it seems to violate the qECT thesis. Therefore Aaronson, who is a proponent of the qECT thesis argued that  there must be  something wrong in my assumptions.

There are a number of possible conclusions which I list here:

\begin{enumerate}
\item The connection between complexity and volume may be wrong. Perhaps there is another less subtle, more easily measured or calculated quantity in the boundary theory, that keeps track of wormhole growth. Bouland, Fefferman, and Vazirani called such a quantity ``pseudo-complexity."
\item Perhaps the complexity-volume duality is correct but the duality between abnormal complexity growth and firewalls is not  general.
\item The big mess: most states are not geometric. The overwhelming majority of states are just too messed up to have a classical geometry behind the horizon. Even states with maximally growing complexity may have firewalls, and therefore Alice  doesn't learn much when she jumps in.
\item The quantum-Extended Church-Turing thesis is wrong.
\item The qECT thesis described in section \ref{Sec: QECT} requires modification.

\end{enumerate}

\bn

First, item $1$: Is it possible that what I am calling complexity is not complexity but rather some much  more feelable\footnote{The term ``feelable" was coined by Bouland, Fefferman, and Vazirani \cite{Bouland:2019pvu} as a shorthand for efficiently measurable or computable. Differences in complexity are extremely un-feelable.} pseudo-complexity quantity? The quantity dual to volume might not  exactly be complexity, but even if so, it will not help.  Bouland, Fefferman, and Vazirani, without assuming  CV duality have argued that whatever 
the holographic dual of wormhole volume is, it must have 
 properties very similar to complexity. In particular BFV argue that it can only be determined on exponential time scales\footnote{The argument is not rigorous but it follows from the assumption that distinguishing a  family of "pseudorandom quantum
states"  from truly random states cannot be done efficiently. The opposite would be surprising. }. This rules out  pseudo-complexity as a resolution of the problem.
 
Item $2$ is more difficult to rule out but I see no evidence for it. The work of Zhao \cite{Zhao:2017iul} suggests that the connection between the rate of volume growth and the existence of measurable effects behind the horizon is  robust. 	

Item 3 maintains that the existence of geometry is a very rare thing and that even states with increasing complexity may be non-geometric. It is true that almost all states are maximally complex with complexity of order $\exp S.$ We know very little about the geometry of such states, but thus far we have only been considering states of low complexity. The question of what happens at exponential time is very interesting but irrelevant.

There is no evidence that low complexity states are messed up; to the contrary, attempts to mess up the interiors of black holes generally fail. An example is the attempt to mess up the interior of a two-sided black hole by repeatedly  throwing in many shock waves either in or out of time order \cite{Shenker:2013pqa}\cite{Shenker:2013yza}. In fact what happens is that instead of the interior becoming a non-geometric mess, the Penrose diagram gets wider making more room, and diluting the shock waves. 

\bn

Let us turn to item $4.$ How likely is it that the qECT thesis is wrong in a gravitational setting? That may depend on exactly how it is formulated. 
In the case at hand, the calculation referred to in the definition is: Given a quantum state, determine the rate of complexity growth; in particular is complexity increasing or decreasing? If Alice can accomplish this in polynomial time then we may claim that the qECT thesis has been violated and therefore cannot be generally correct.
In fact, by entering the black hole Alice  can do exactly that.

Can the qECT thesis be rescued? The earlier version:

\bn

\it  Any calculation that cannot be done efficiently  by a quantum Turing machine (or quantum circuit), cannot be done efficiently by any physical system consistent with the laws of physics. \rm

\bn
is too strong. We might entertain the following modification:

\bn
\it  Any calculation that cannot be done efficiently  by a quantum Turing machine (or quantum circuit), cannot be done efficiently by any physical system which remains  able to communicate with the holographic boundary of space. \rm

\bn

In other words the thesis applies only to physical systems which remain outside the horizon.
Since observers who have passed the horizon cannot communicate their results to boundary observers, the fact that they may learn the properties of complexity growth shortly after entering the black hole does not count as a violation of the qECT thesis. This would be a satisfying resolution of the dilemma in which the horizon has a new information-theoretic role as the protector of the qECT thesis. 

However, that is a bit too fast. We know that it is possible in quantum gravity for information to be retrieved from behind the horizon; for example using quantum teleportation protocols, not to transmit information but to probe the interior of a wormhole. If this were possible efficiently we would have a genuine contradiction with the qECT thesis.

Thus the qECT thesis makes a non-trivial prediction: 

\bn

\it Any apparent violation of the qECT thesis behind the horizon of a black hole cannot be communicated to the black hole exterior efficiently.
\rm

\bn

In particular, information about the volume of a wormhole and the rate of change of the volume, gained behind the horizon by Alice,  cannot be retrieved  outside the horizon in less than exponential time. 

\section{Conclusion}

Granted some technical assumptions, the quantum-extended Church-Turing thesis, as stated at the beginning of this paper, is wrong. By crossing a black hole horizon Alice  can very quickly determine whether the rate of complexification is abnormally slow or negative. Alice  has therefore made a calculation in a time that would be forbidden by the qETC thesis.

The qETC can be rescued if we  require that it only  hold for observers who remain in communication with the holographic boundary. This would mean that all ordinary quantum experiments done on the boundary CFT  will respect the thesis.
From a computer science point of view this is all we care about. A technician operating and observing a quantum computer is like a boundary observer with control over the CFT, not an observer who falls through the horizon. 

 But from a fundamental physics point of view we learn something interesting about the black hole interior; namely, what is simple and physically  feelable in the bulk region  behind the horizon, may be very complex  and un-feelable when viewed from the boundary point of view. This of course is also the point  made in \cite{Bouland:2019pvu}: the bulk-boundary dictionary is highly complex.
 
 This fact has implications for theory if not for experiment. Attempts to find an explicit encoding of the interior geometry of a black hole in terms of boundary degrees of freedom are probably doomed, not because such an encoding is impossible, but because it would require exponential resources.

\section*{Acknowledgements}

I thank  Scott Aaronson and Adam Bouland for  discussions which greatly influenced  my thinking about the issues in this paper. Discussions with Adam Brown and Henry Lin were also valuable.

\end{document}